\documentclass[twoside,12pt]{article}
\usepackage{epsfig}

\newcommand{\be}{\begin{equation}}
\newcommand{\ee}{\end{equation}}
\newcommand{\bea}{\begin{eqnarray}}
\newcommand{\eea}{\end{eqnarray}}

\newcommand{\etal}{\emph{et al.}}

\newcommand{\nue}{\mbox{$\nu_{\rm e}$}}
\newcommand{\numu}{\mbox{$\nu_{\mu}$}}

\newcommand{\nuord}{\mbox{$\nu_{\rm ord}$}}
\newcommand{\nuextra}{\mbox{$\nu_{\rm extra}$}}
\newcommand{\costh}{\mbox{$\cos \theta$}}
\newcommand{\sinth}{\mbox{$\sin \theta$}}

\begin{document}

\title{ \vspace{1cm} Neutrino oscillations with a polarized laser beam:\\
an analogical demonstration experiment}
\author{C.\ Weinheimer\\
Institut f\"ur Kernphysik, Westf\"alische Wilhelms-Universit\"at M\"unster, Germany\\
Email: weinheimer@uni-muenster.de}
\maketitle
\begin{abstract} 
The underlying physics of neutrino oscillation in vacuum can be demonstrated by an optical analogical experiment.
Two different neutrino flavors are represented by two polarization states of a laser beam, whereas the different phase
propagation in vacuum is mimicked by the propagation difference of an ordinary and an extraordinary beam in a birefringent crystal.
This allows us to demonstrate neutrino oscillation by optical methods 
in a fully microscopic way at the particle level.
The description of both realizations of oscillation is also mathematically identical.
In our demonstration experiment we can vary the oscillation parameters 
such as propagation length $L$ and mixing angle $\theta$.
\end{abstract}
\section{Introduction}
Since the first clear evidence in 1998 from the Super Kamiokande experiment \cite{superkamiokande_atmos} 
we have known that neutrinos from one flavor ({\it e.g.} electron, muon
or tau neutrinos) can change into another flavor during flight and vice versa.
Such ``neutrino oscillation'' is caused by neutrino flavor mixing and non-zero neutrino masses. 
Meanwhile neutrino oscillation has been confirmed by many experiments using 
atmospheric, solar, reactor and accelerator neutrinos and its analysis has been 
expanded to the oscillation of three neutrino flavors including matter effects.  

It is very intriguing for students to learn, that, for example,
electron neutrinos created in the core of the sun by
nuclear fusion are detected on earth as muon or tau neutrinos. 
Additional fascination comes from the fact that most experts within the community consider the
existence of neutrino oscillation as first evidence for physics beyond the Standard Model of particle physics, 
since it is very unnatural to explain 
why neutrinos are at least six orders of magnitude lighter than
the charged fundamental fermions, by very different Yukawa couplings of neutrinos to the Higgs boson. 
It would be more natural that neutrinos are their own antiparticles and
that they possess additional, lepton-number violating Majorana mass terms, which could explain the smallness of neutrino masses
in a much more natural way. Consequently, non-zero masses of neutrinos and their mass pattern are very important to differentiate between 
models of particle physics beyond the Standard Model \cite{lindner}. 
Looking into the universe, relic neutrinos left over from the Big Bang are 
the only dark matter component which we are sure about. Therefore, the neutrino mass values are important for understanding the evolution 
of the universe and, especially, structure formation \cite{hannestad}. 
In addition, neutrino mixing and oscillation may play a decisive role in supernova explosions \cite{raffelt}. 

Therefore, neutrino oscillation is not only fascinating, it is of such great importance for nuclear and particle physics as well as
for astrophysics and cosmology, and its evidence so compelling,
that it has become standard textbook knowledge for third-year or fourth-year
students of physics.
Usually, neutrino oscillation is taught when the students already have 
reasonable knowledge of quantum mechanics and when they understand the
concepts of non-communicating operators and different eigenvectors. This teaching is purely mathematical and is underlined by published data from various neutrino
oscillation experiments. To demonstrate or to measure neutrino oscillation in a laboratory class experiment is totally impossible
due to the small neutrino cross sections and the long oscillation lengths. 

On the other hand, the physics of neutrino oscillation can be apprehended by understanding mixed states and simple oscillation or wave theory. 
Quite well known are the pendulum demonstration experiments with coupled double and triple pendula \cite{kobel}. Here the two fundamental ingredients
of neutrino oscillation -- mixing of states of different bases and time evaluation of a coherent superposition of different states -- can be caught and even touched by hand. 
On the other hand, objects such as oscillation modes of macroscopic pendula are not obviously representing the behavior of microscopic quantum states or particles. 
Therefore, we would like to present a microscopic demonstration experiment of the physics of neutrino oscillation which is based on two states of polarized light
going through a birefringent crystal. We will show that this experiment is fully analogical to two-flavor neutrino oscillation in vacuum, even the mathematical calculation
is fully identical. Another advantage is that the students are already used to one of the important ingredients -- coherent superposition -- by optical interference experiments
like the double slit.
The experiment is rather simple and can be
taught to first-year or second-year students in an optics class. 
It might also be useful for outreach or even for 
students in a high-level physics class at high school.

This article is structured as follows: In section 2 we briefly recapitulate two flavor neutrino oscillation in vacuum.
In section 3 we present the idea of the demonstration experiment and calculate the transition probability, which we compare to section 2.
In section 4 we present the real experiment and its data. The conclusions are given in section 5.

\section{Two flavor neutrino vacuum oscillation}
Neutrino oscillation requires neutrino mixing: The neutrino flavor eigenstates ($\nue, \numu, \nu_\tau$) are not identical
to the neutrino mass eigenstates ($\nu_1, \nu_2, \nu_3$), but are connected 
via an unitary matrix $U$ similar to the quark case.
We restrict ourselves to the two-flavor mixing case and neglect matter effects (``vacuum oscillation''), 
which is sufficient to describe the underlying 
physics\footnote{In a three-flavor mixing scenario there appears a CP-violating phase in the mixing matrix, which possibly creates differences between
the oscillation of neutrinos and antineutrinos.}.:

\begin{equation}
\left( 
  \begin{array}{c}
     {\nue } \\ {\numu }
  \end{array} 
\right)
=
\underbrace{  \left(
  \begin{array}{cc}
    \costh & -\sinth \\ \sinth & \costh
  \end{array} 
\right)}_{U}
\left( 
  \begin{array}{c}
     \nu_1 \\ \nu_2
  \end{array} 
\right)
\qquad
\left( 
  \begin{array}{c}
     \nu_1 \\ \nu_2
  \end{array} 
\right)
=
\underbrace{  \left(
  \begin{array}{cc}
    \costh & \sinth \\ -\sinth & \costh
  \end{array} 
\right)}_{U^{-1}}
\left( 
  \begin{array}{c}
     {\nue } \\ {\numu }
  \end{array} 
\right) \label{eq::nu_mixing}
\end{equation}

\begin{figure}[t!]
\begin{center}
  \begin{minipage}{0.58\textwidth}
    \epsfig{file=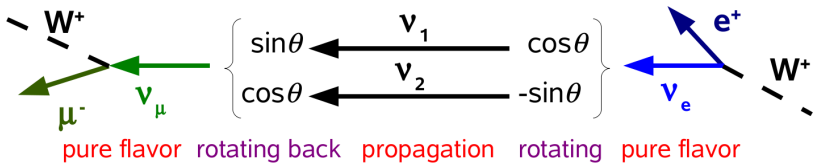,width = \textwidth}
  \end{minipage}
  \hfill
  \begin{minipage}{0.40\textwidth}
    \caption{Schematic picture of two-flavor neutrino vacuum oscillation.
      The weak charge current interactions
      via $W$ boson exchange define the neutrino flavor through the charged lepton, whereas the propagation
      takes place as coherent superposition of mass eigenstates.  \label{fig::nu_osc}}
  \end{minipage}
\end{center}
\end{figure}

We can read from fig. \ref{fig::nu_osc} the probability $P({\nue } \rightarrow {\numu })$ that a defined flavor \nue\ 
is detected as flavor \numu\ after the neutrino
has propagated over a distance $L$ as a coherent 
superposition of mass eigenstates $\nu_1$ and $\nu_2$. It is most convenient to
use two-dimensional notations and $\hbar = 1 = c$ (Please see appendix 3 for
the detailed derivation of equation (\ref{eq::osc_formula}).):

\begin{eqnarray}
P({\nue } \rightarrow {\numu })
=& 
\left| 
  \left( 
    \begin{array}{c}
       0 \\ 1 
    \end{array} 
  \right) 
  U 
  \left(
     \begin{array}{cc} 
        e^{-i E_1 t} & 0 \\ 0 & e^{-i E_2 t} 
     \end{array}
  \right)
  U^{-1}
  \left( 
    \begin{array}{c} 
       1 \\ 0 
    \end{array} 
  \right) 
\right| ^2  \\
=& 
\sin^2 (2\theta) \sin^2 \left(\frac{\Delta m^2 L}{4E} \right) \label{eq::osc_formula}\\
= & \sin^2 (2\theta) \sin^2 \left( \pi \frac{L}{\lambda_{\rm osc} } \right) 
\quad {\rm with} \quad \lambda_{\rm osc} = \frac{4\pi E}{\Delta m^2}
\end{eqnarray}

\section{The idea of the optical analogical experiment for neutrino oscillation}
The idea of our demonstration experiment is based on the two state system of polarized photons. We can define a pure ``flavor'' state
by sending a laser beam through a linear polarizer. 
After the propagation  we measure the intensity in the same ``flavor'' 
state by a detector behind a linear polarizer, called analyzer,
parallel to the first polarizer or in the other ``flavor'' state by an analyzer
which is oriented orthogonal to the first polarizer.

\begin{figure}[b!]
\begin{center}
    \epsfig{file=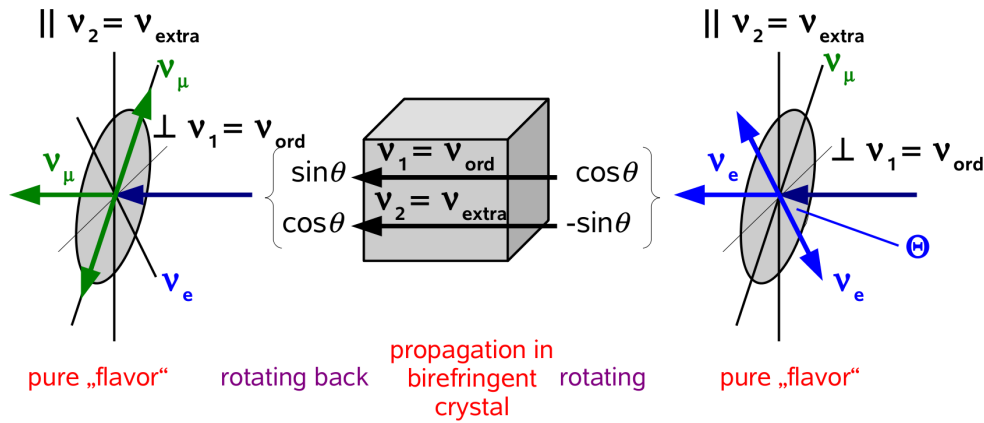,width = 0.7\textwidth}
    \caption{Schematic picture of the optical analogical experiment to demonstrate neutrino oscillation. A ``flavor'' state defined by the polarizer is propagating
       as a coherent superposition of ordinary and extraordinary states through a birefringent crystal, whose optical axis is vertical. The ``flavor'' state is checked by
       an analyzer. \label{fig::nu_osc_pol}}
\end{center}
\end{figure}
To obtain the oscillation effect 
we need a phase propagation difference for another set of polarized states, which is rotated with respect to the ``flavor'' base.
This requirement is realized by a birefringent crystal, whose optical axis lies in the plane perpendicular 
to the beam direction, but which is rotated with respect to the direction of both ``flavor'' states. Photons whose polarization direction is perpendicular to the optical axis of the 
birefringent crystal propagate as an ``ordinary'' beam with the refraction index $n_{\rm ord}$. Photons whose polarization vector is -- at least partly -- 
parallel
to the optical axis propagate as an ``extraordinary'' beam with refraction index $n_{\rm extra}$. The value of $n_{\rm extra}$  
depends on the angle between the polarization vector and the 
optical axis and deviates maximally from $n_{\rm ord}$ when the polarization vector is parallel to the optical axis. 
Usually birefringence is demonstrated by the splitting of a beam due to the different indices of refraction, {\it e.g.} double image using a calcite
crystal. If the entrance plane of the crystal contains the optical axis and is perpendicular to the light beam then the birefringence yields 
no beam splitting but a different phase propagation of the ordinary and the extraordinary beam. Thus a  ``$\lambda/4$'' plate can turn linear polarized light into circular polarized light and vice versa.

The idea of our neutrino oscillation analogical experiment with polarized light is illustrated in figure \ref{fig::nu_osc_pol}. The two ``flavor'' states are defined by the directions
of linear polarizers, which enclose angles of $\theta$ and $90^o - \theta$ with respect to the horizontal. A birefringent crystal is placed with its entrance plane
perpendicular to the laser beam. Its optical axis is exactly vertical.

To calculate the light propagation through the birefringent crystal we have to consider a pure ``flavor'' state (``\nue'' or \numu'')
defined by the orientation of the polarizer (see fig. \ref{fig::nu_osc_pol})
as a superposition of the ordinary beam \nuord\ and the extraordinary beam 
\nuextra . The transformation matrices are exactly the same as for neutrinos (equation (\ref{eq::nu_mixing})):

\begin{equation}
\left( 
  \begin{array}{c}
     {``\nue ''} \\ {``\numu ``}
  \end{array} 
\right)
=
\underbrace{  \left(
  \begin{array}{cc}
    \costh & -\sinth \\ \sinth & \costh
  \end{array} 
\right)}_{U}
\left( 
  \begin{array}{c}
     \nu_{\rm ord} \\ \nu_{\rm extra}
  \end{array} 
\right)
\qquad
\left( 
  \begin{array}{c}
    \nu_{\rm ord} \\ \nu_{\rm extra}
  \end{array} 
\right)
=
\underbrace{  \left(
  \begin{array}{cc}
    \costh & \sinth \\ -\sinth & \costh
  \end{array} 
\right)}_{U^{-1}}
\left( 
  \begin{array}{c}
     {``\nue''} \\ {``\numu''}
  \end{array} 
\right)
\end{equation}

Then the probability $P({`` \nue `` } \rightarrow {`` \numu ``})$
of a transition from flavor ``\nue `` to ``\numu'' after propagating through
a birefringent crystal of thickness $L$ can be calculated:

\begin{eqnarray} 
P({`` \nue `` } \rightarrow {`` \numu ``}) 
=& 
\left| 
  \left( 
    \begin{array}{c}
       0 \\ 1 
    \end{array} 
  \right) 
  U 
  \left(
     \begin{array}{cc} 
        e^{- i 2\pi L n_{\rm ord} /\lambda} & 0 \\ 0 & e^{- i 2 \pi L n_{\rm extra} / \lambda } 
     \end{array}
  \right)
  U^{-1}
  \left( 
    \begin{array}{c} 
       1 \\ 0 
    \end{array} 
  \right) 
\right| ^2  \label{eq::pol_osc_ansatz}\\
=& 
\sin^2 (2\theta) \sin^2 \left(\frac{\Delta n L}{\lambda / \pi} \right) \label{eq::pol_osc_formula}\\
= & \sin^2 (2\theta) \sin^2 \left( \pi \frac{L}{\lambda_{\rm osc} } \right) 
\quad {\rm with} \quad \lambda_{\rm osc} = \frac{\lambda}{\Delta n} \label{eq::pol_osc_length}
\end{eqnarray}
The transition from equations  (\ref{eq::pol_osc_ansatz}) to (\ref{eq::pol_osc_formula}) requires the same algebraic steps
as for the calculation of neutrino oscillation (See appendices 3 and 4 for details.).
Equation (\ref{eq::pol_osc_formula}) is mathematically identical to the oscillation formula of two-flavor neutrino oscillation
in vacuum (\ref{eq::osc_formula}). 
For didactical reasons we might even 
identify variables of neutrino flavor oscillations with 
the corresponding variables of the demonstration experiment.
\begin{displaymath}
  L \equiv L \qquad  \Delta m^2 \equiv \Delta n \qquad E \equiv \frac{\lambda}{4\pi}
\end{displaymath}

\section{The demonstration experiment and data}

\begin{figure}[t!]
\begin{center}
    \epsfig{file=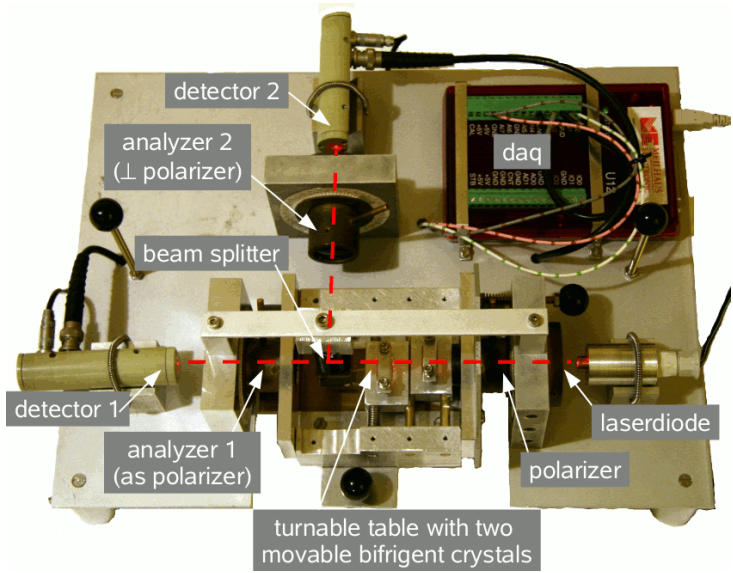,width = 0.52\textwidth}
    \caption{Picture of the analogical experiment to demonstrate the physics of neutrino oscillation. The beam of a red laser pointer is projected on a defined ``flavor'' state
      by a polarizer before it enters a doublet of birefringent crystals, which allow us to vary the length of the geometrical path through the birefringent material $L$. 
      These crystals are mounted on a rotatable optical table to allow different mixing angles $\theta$. 
      Then the beam is split by a beam splitter and sent through two orthogonal analyzers to photodiodes in order to record at the same time 
     the intensity in both ``flavor'' states.
      The current of the photodiodes is amplified,converted into a voltage and digitized with a data acquisition board (``daq''), which is
     connected via a USB link to a computer running LabView.
     \label{fig::nu_osc_setup}}
\end{center}
\end{figure}

Figure \ref{fig::nu_osc_setup} shows the demonstration experiment, which is mounted on an aluminum sheet to be transported in a small suitcase. 
We use lithium tetra borate crystals (Li$_2$B$_4$O$_7$, LTB) of dimensions
$15 \times 10 \times 10$~mm$^3$ in the [100], [010] and [001] directions. 
The optical axis lies in the [001] direction and is oriented vertically in our setup.
The two indices of refraction are
$n_{\rm ord} = 1.552$ and $n_{\rm extra}=1.609$ (polarization vector parallel to the optical axis) 
at the wavelength of our laser pointer $\lambda = 633$~nm.

\begin{figure}[h!]
\begin{center}
  \begin{minipage}{0.12\textwidth}
    \epsfig{file=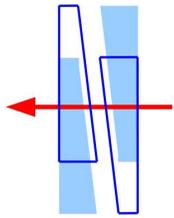,width = \textwidth}
  \end{minipage}
  \hfill
  \begin{minipage}{0.86\textwidth}
    \caption{Schematic picture of variation of the length through the birefringent material $L$ by moving the two slightly trapezoidal-shaped crystals against each other.
      Each crystal moves from one position (solid trapezium) to the other (open trapezium) by 5.5~mm transversal to the laser beam direction (arrow) by a manipulator, which is
      connected to a linear potentiometer in order to read out the crystals` positions by the computer via the data acquisition board. \label{fig::moving_crystals}}
  \end{minipage}
\end{center}
\end{figure}

To allow us to vary the geometrical
path through the birefringent material $L$ we are using a doublet of crystals, which have a slightly trapezoidal shape with a non-zero wedge angle.
By moving them against each other, $L$ is changed (see fig. \ref{fig::moving_crystals}).  The surfaces
normal to the [010] direction have been polished and possess a wedge angle of $0.2^o$ against each other.  The variation of the transversal position of the crystals by 5.5~mm
each corresponds to a change of the geometrical path through the birefringent crystals of $\Delta L = 38~\mu$m (see fig. \ref{fig::moving_crystals}), which corresponds
to about 3.5 oscillations lengths 
$\lambda_{\rm osc} = \lambda/\Delta n = 11~\mu$m 
(see equation (\ref{eq::pol_osc_length})). 
To allow us to change 
the mixing angle $\theta$ the optical table with the birefringent crystals can be rotated with respect to the position of the polarizer\footnote{Of course it looks simpler
to rotate the polarizer instead, but this would require that we would have to rotate both analyzers exactly by exactly the same angle, which is more difficult to realize 
in a demonstration experiment.} (See pictures in 
appendix 2.).

When the geometrical length is varied by the manipulator, the positions of the crystals and the light intensities at both detectors are recorded via a LabView-based software\footnote{Laboratory Virtual Instrument Engineering Workbench by the company National Instruments}
on a computer and visualized directly on the screen or a beamer (see figure \ref{fig::measurements}). Some exemplary data are shown in appendix 1.

\begin{figure}[t!]
\begin{center}
    \epsfig{file=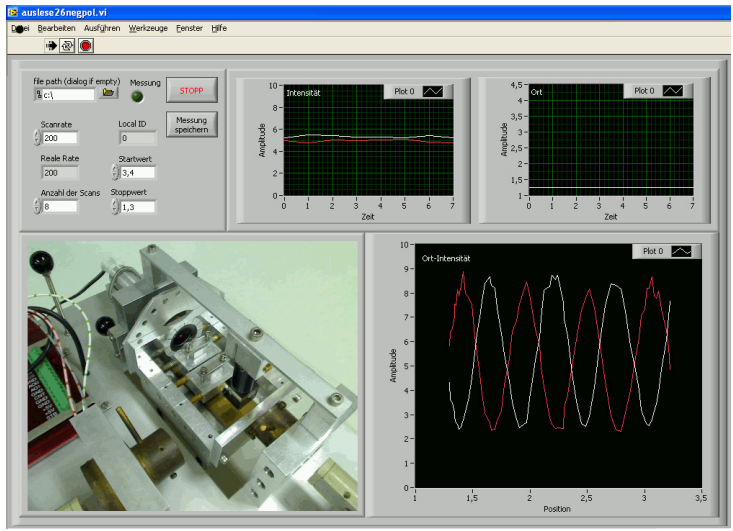,width = 0.52\textwidth}
    \caption{Screen shot of the LabView program illustrating the status of the experiment on different sub-windows. 
      Upper left: control parameter panel; 
      Upper middle: actual intensities measured in both detectors;
      Upper right: position of the moving crystals;
      Lower left: live picture taken by a webcam to illustrate the movement of the crystals and to show the rotation angle of the optical desk; Lower right:
      online diagram of the last measured intensities in both detectors versus the change of the propagation length in the birefringent
      crystal $L$, {\it i.e.} the oscillation plot.
     \label{fig::measurements}}
\end{center}
\end{figure}

\section{Conclusions}
We have shown that the underlying physics of neutrino oscillation 
--  the mixing of states of different bases and time evaluation of a coherent superposition of different states --
can be demonstrated by sending a polarized laser beam through a birefringent crystal
and using the different states of polarized light defined by a polarizer and within a birefringent crystal. 
The derivation of the two-flavor neutrino oscillation formula and of our oscillation formula of polarized light 
correspond one-by-one.
By varying the geometrical path through the birefringent crystals and by varying the mixing angle between the 
external polarization states (``flavor'' states) and the states defined by the propagation in the birefringent crystals
(the ordinary and extraordinary beams: ``mass'' states) we can investigate all details of the two-flavor neutrino oscillation formula
in vacuum. 
Our realization is a full microscopic analogical experiment at the particle level, since -- given single-photon sensitive detectors -- 
one could reduce the power of the laser light source such that only one photon at a time is flying through the setup.

This experiment has been set up on a small aluminum table in order to transport it to lecture rooms for outreach talks. 
The experiment can be visualized on a beamer via a laptop. We have checked, that the whole experiment can also be set up by using standard laboratory class equipment 
of a high school with the exception of the two trapezoidal-shaped birefringent crystals\footnote{The experiment
would work as well with only one birefringent crystal, but the laser beam would be bent a little bit.}, which need to be bought from an optics company.

\section*{Acknowledgment}
The author gratefully acknowledges the great technical support by G. Hackmann and H. Baumeister in constructing and setting up the experiment. 
The author would like to thank P. Boschan for many helpful discussions, C. Kraus for helping to set up a very first version of this demonstration experiment,
T. Bode for testing a prototype version of this experiment, A. Bakenecker for testing the present experiment 
and A. Peter from the Hungarian Academy of Sciences, Budapest for providing us with the birefringent crystals.

\newpage

\section*{Appendix 1: Exemplary data}
\begin{figure}[h!]
\begin{center}
    \epsfig{file=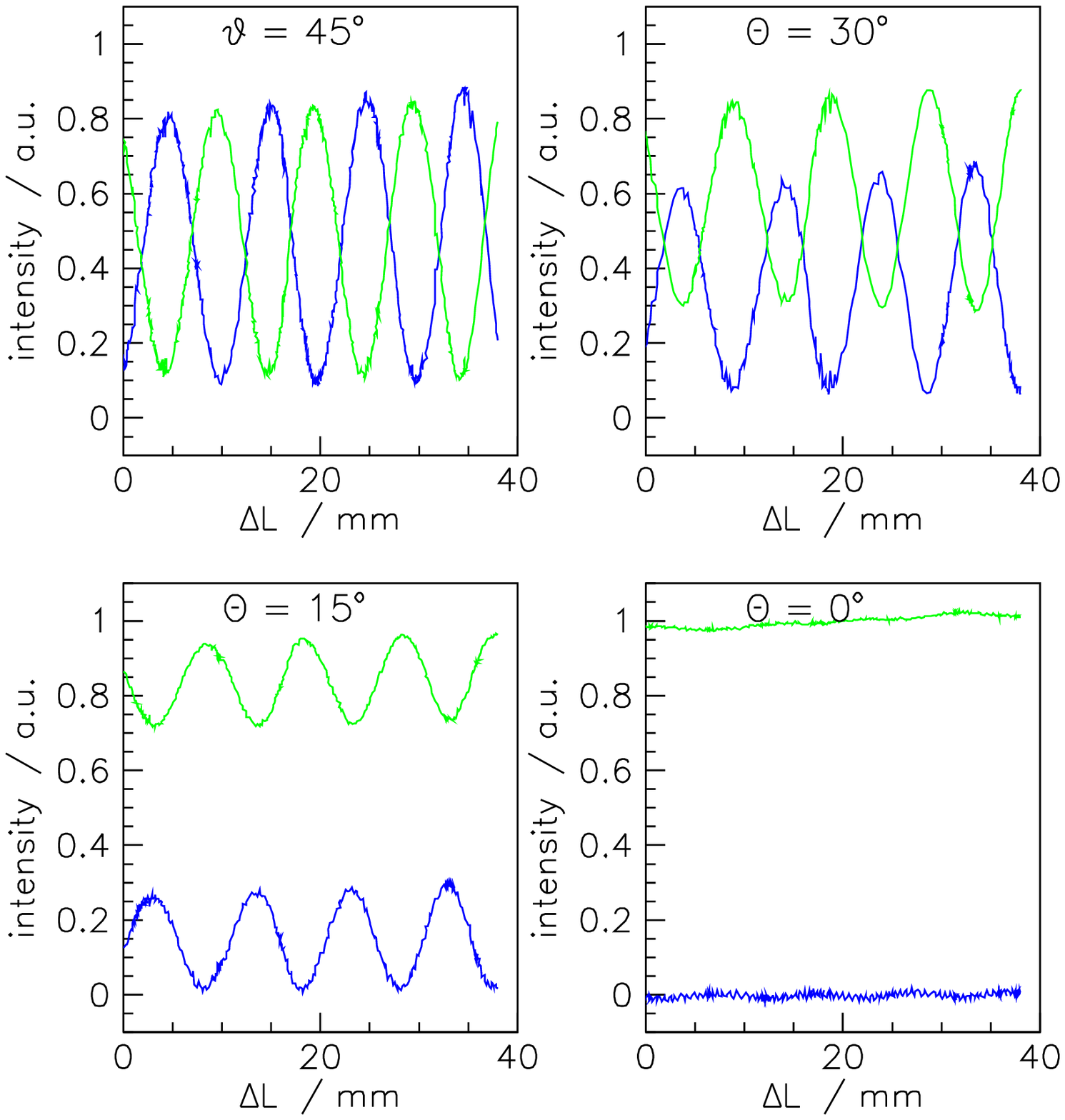, width = 0.49\textwidth}    
    \epsfig{file=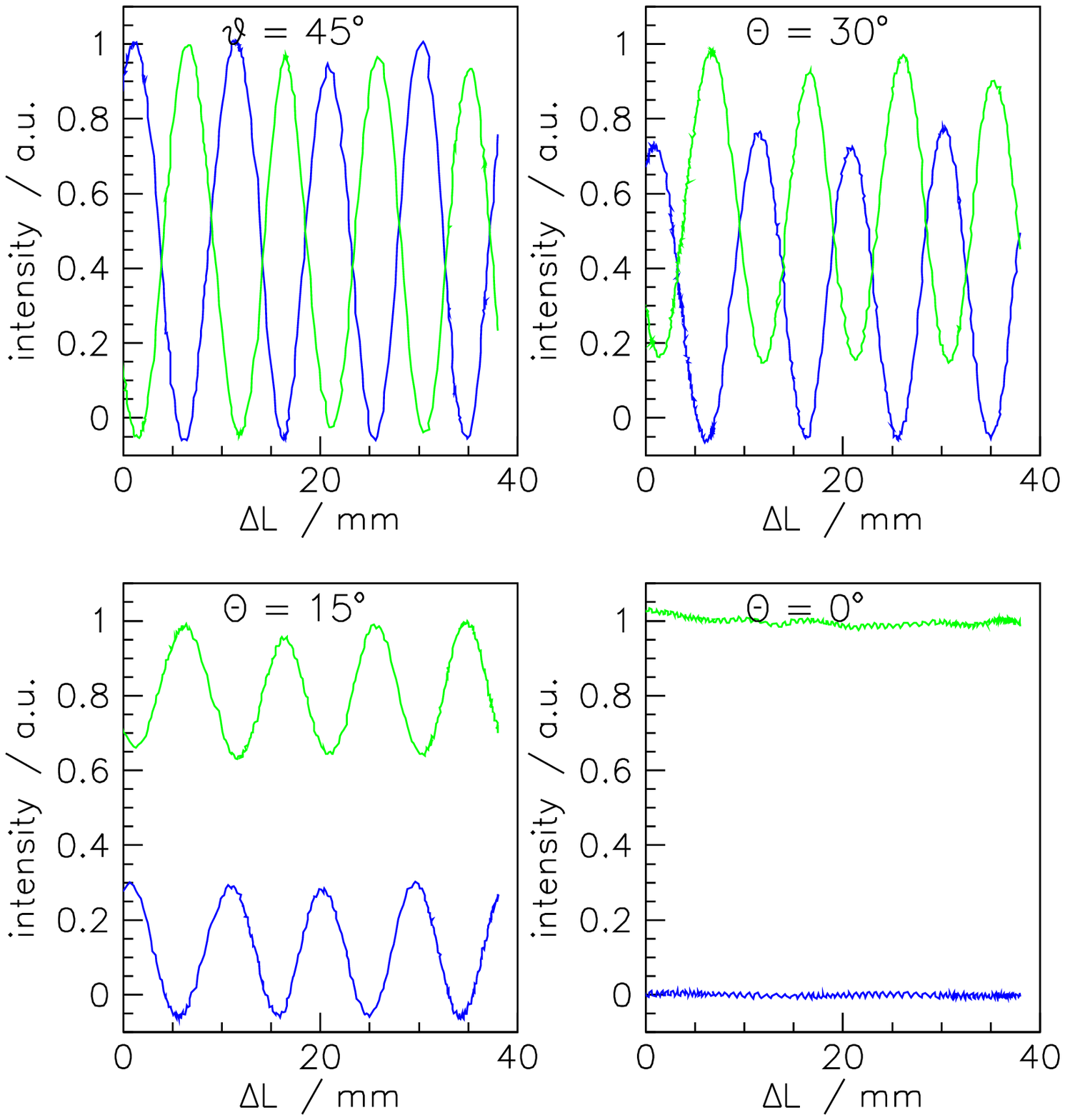, width = 0.49\textwidth}
    \caption{Intensities measured by the detector 1   (green) 
         and detector 2 (blue) for different rotation angles 
         ({\it i.e.} mixing angles $\theta$) 
         of the rotatable table housing the
         birefringent crystals. Left 4 plots: measured with the laser diode; Right 4 plots:
         measured with a HeNe laser. The data follow 
         the behavior expected from equation (\ref{eq::osc_formula})
         as function of the propagation length $L$ and mixing angle
         $\theta$. The quality of our laser pointer and especially
         its spectral widths yields some out-washing effect
         of the oscillation pattern. This is especially visible
         for the data at $\theta=45^o$ and at $\theta=30^o$ by the
         fact, that they do not range in intensity from 0 to 1.
         Using a HeNe laser (plots on the right half) this out-washing effect does not exist, 
         proving that it is caused by the laser pointer and not by 
         the birefringent crystals or another part of the setup.}
  \label{fig::laserdiode_data}
\end{center}
\end{figure}

\section*{Appendix 2: More pictures of the setup}

\begin{figure}[h!]
\begin{center}
    \epsfig{file=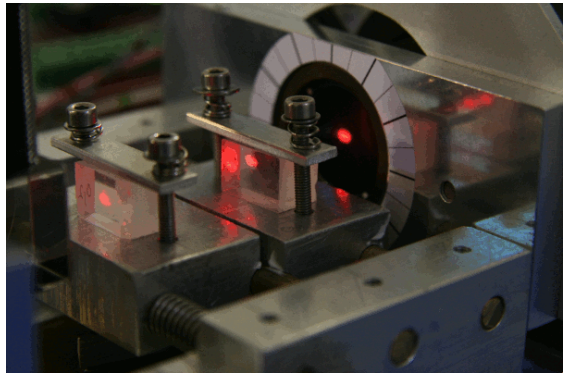,width = 0.4\textwidth}
    \caption{Close-up view of the table with the two birefringent crystals.}
\end{center}
\end{figure}

\begin{figure}[h!]
\begin{center}
    \epsfig{file=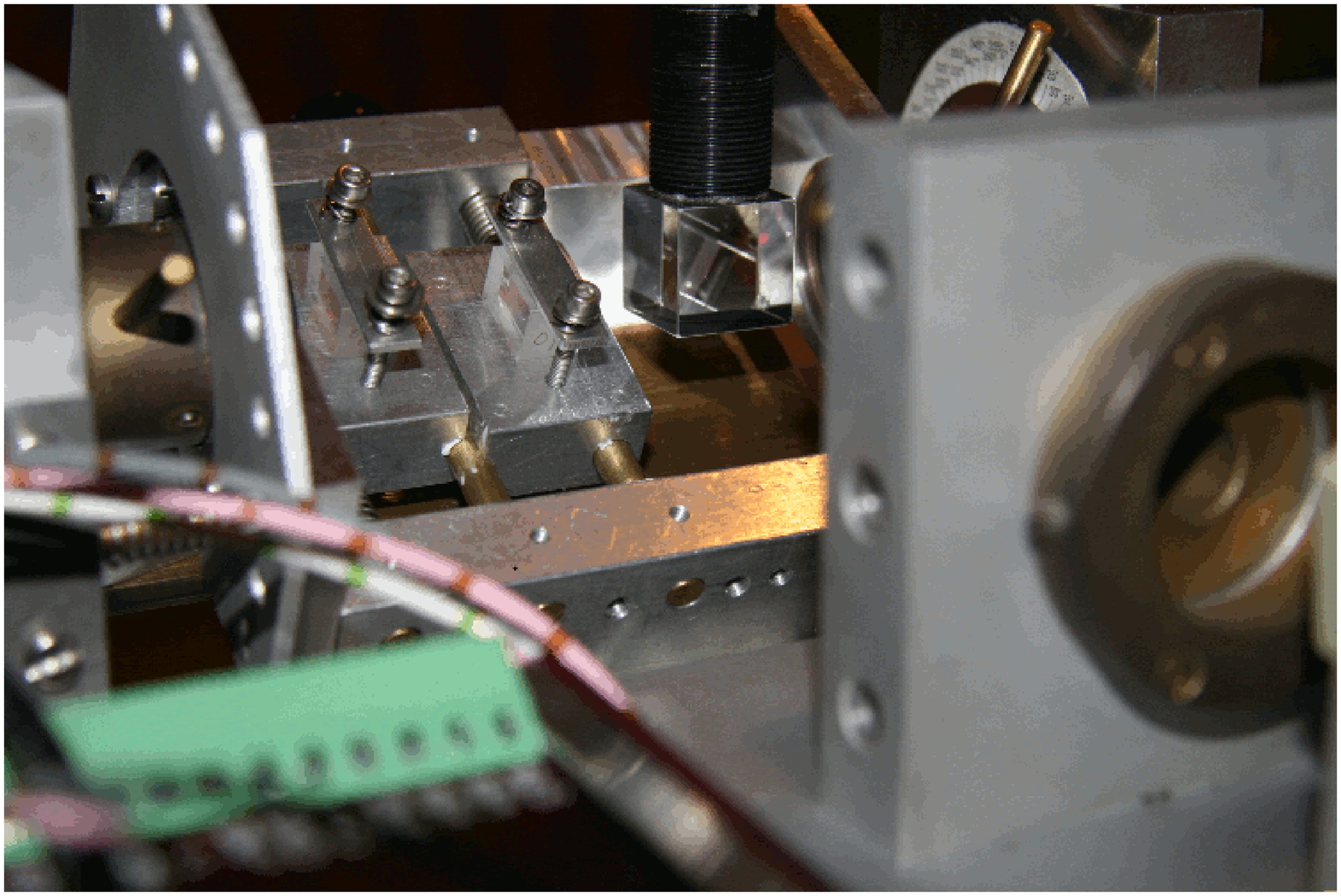,width = 0.4\textwidth} \hspace*{1cm}
    \epsfig{file=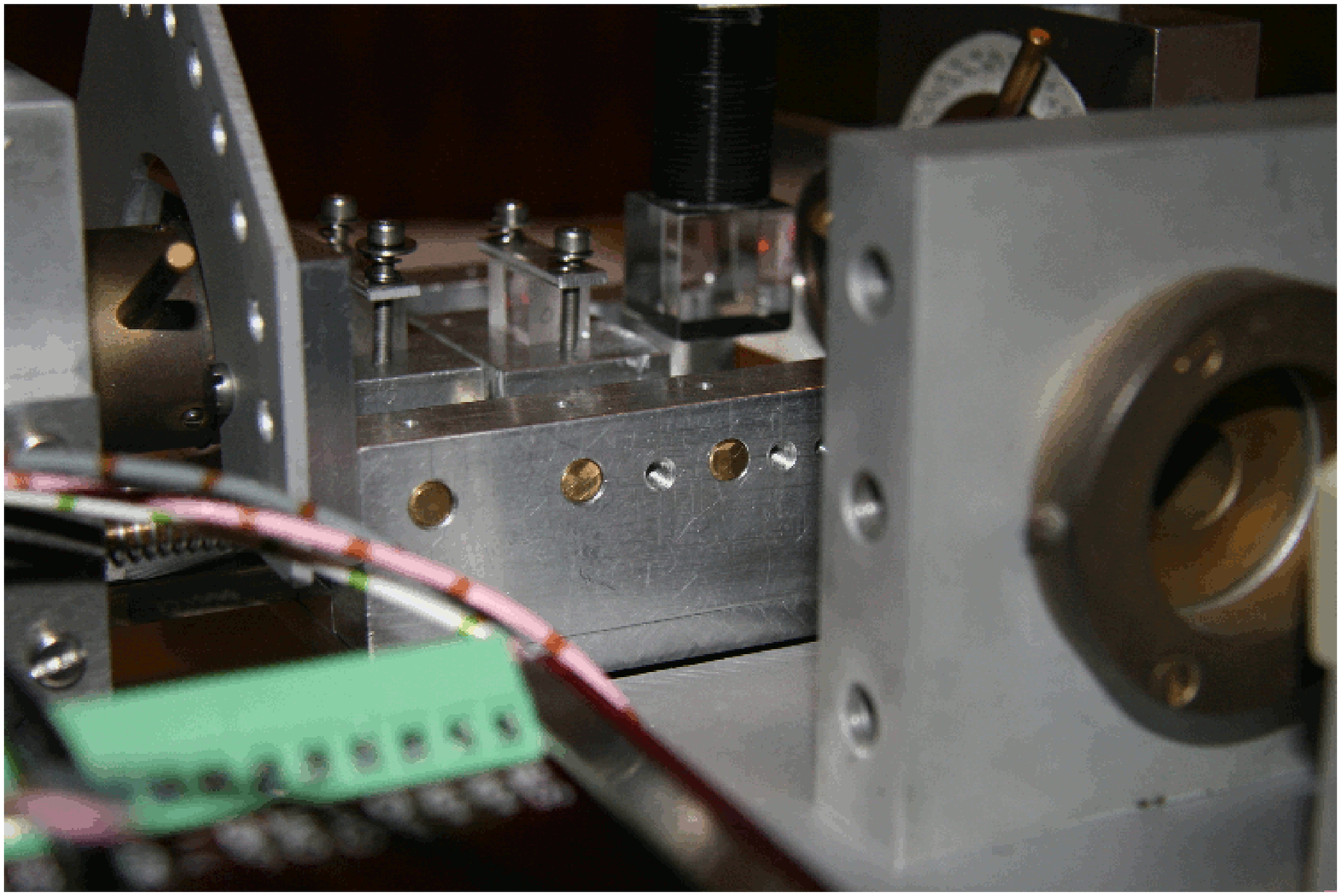,width = 0.4\textwidth}
    \caption{The rotatable table with the two birefringent 
    crystals. Left: rotated; Right: standard orientation 
    (resulting in $\theta=45^o$).}
\end{center}
\end{figure}

\section*{Appendix 3: Derivation of equation (\ref{eq::osc_formula})}

\begin{eqnarray}
P({\nue } \rightarrow {\numu })
=& 
\left| 
  \left( 
    \begin{array}{c}
       0 \\ 1 
    \end{array} 
  \right) 
  U 
  \left(
     \begin{array}{cc} 
        e^{-i E_1 t} & 0 \\ 0 & e^{-i E_2 t} 
     \end{array}
  \right)
  U^{-1}
  \left( 
    \begin{array}{c} 
       1 \\ 0 
    \end{array} 
  \right) 
\right| ^2  \\
=& 
\left| 
  \left( 
    \begin{array}{c}
       0 \\ 1 
    \end{array} 
  \right) 
  \left( 
    \begin{array}{cc} 
      \costh & -\sinth \\ \sinth & \costh 
    \end{array} 
  \right) 
  \left(
     \begin{array}{cc} 
        e^{-i E_1 t} & 0 \\ 0 & e^{-i E_2 t} 
     \end{array}
  \right)
  \left(
    \begin{array}{cc} 
       \costh & \sinth \\ -\sinth & \costh
    \end{array}
  \right) 
  \left( 
    \begin{array}{c} 
       1 \\ 0 
    \end{array} 
  \right) 
\right| ^2  \\
=& 
\left| 
  \left( 
    \begin{array}{c}
       0 \\ 1 
    \end{array} 
  \right) 
  \left( 
    \begin{array}{cc} 
      \costh & -\sinth \\ \sinth & \costh 
    \end{array} 
  \right) 
  e^{-i E_1 t} 
  \left( 
     \begin{array}{cc} 
        1 & 0 \\ 0 & e^{-i (E_2-E_1)t} 
      \end{array} 
  \right) 
  \left( 
    \begin{array}{c} 
      \costh \\ -\sinth 
    \end{array} 
  \right) 
\right| ^2 \label{eq::e2me1}\\
=& 
\left| 
  \left( 
    \begin{array}{c}
       0 \\ 1 
    \end{array} 
  \right) 
  \left( 
    \begin{array}{cc} 
      \costh & -\sinth \\ \sinth & \costh 
    \end{array} 
  \right) 
  \left( 
    \begin{array}{cc} 
        1 & 0 \\ 0 & e^{-i \frac{\Delta m^2 L}{2E}} 
    \end{array} 
  \right)
  \left( 
    \begin{array}{c} 
      \costh \\ -\sinth 
    \end{array} 
  \right) 
\right| ^2 \label{eq::deltam2}\\
=& 
\left| 
  \left( 
    \begin{array}{c}
       0 \\ 1 
    \end{array} 
  \right) 
  \left( 
    \begin{array}{cc} 
      \costh & -\sinth \\ \sinth & \costh 
    \end{array} 
  \right) 
  \left( 
    \begin{array}{c} 
       \costh \\ -\sinth e^{-i \frac{\Delta m^2 L}{2E}}
    \end{array} 
  \right) 
\right| ^2 \\
= & 
\left|
  \costh \sinth -\costh \sinth e^{-i \frac{\Delta m^2 L}{2E}}
\right| ^2 \\
= & 
\left|
  \costh \sinth \cdot \left( 1 - e^{-i \frac{\Delta m^2 L}{2E}} \right) 
\right|^2 \\
= &
\frac{1}{4} \sin^2 (2\theta) \cdot \left|
 1 - e^{-i \frac{\Delta m^2 L}{2E}} 
\right|^2 \\
= &
\frac{1}{4} \sin^2 (2\theta) \cdot \left(
\left(1 - \cos \left(  \frac{\Delta m^2 L}{2E} \right) \right)^2 
+ \sin^2  \left(  \frac{\Delta m^2 L}{2E} \right) \right)\\
= &
\frac{1}{4} \sin^2 (2\theta) \cdot 2 
\cdot \left(1-\cos \left( \frac{\Delta m^2 L}{2E}\right) \right)\\
= &
\frac{1}{4} \sin^2 (2\theta) \cdot 2 
\cdot \left(1-\cos^2 \left( \frac{\Delta m^2 L}{4E}\right) + \sin^2 \left( \frac{\Delta m^2 L}{4E}\right)
\right)\\
=& 
\sin^2 (2\theta) \sin^2 \left(\frac{\Delta m^2 L}{4E} \right)
\label{eq::osc2_formula} \\
= & \sin^2 (2\theta) \sin^2 \left( \pi \frac{L}{\lambda_{\rm osc} } \right) 
\quad {\rm with} \quad \lambda_{\rm osc} = \frac{4\pi E}{\Delta m^2}
\end{eqnarray}

For the transition from equation (\ref{eq::e2me1}) to equation (\ref{eq::deltam2}) we used the following expansion of the relativistic energy-momentum-relation:
\begin{equation}
  E_i = \sqrt{p_i^2 + m_i^2} = p_i + \frac{m_i^2}{2p_i} \approx E + \frac{m_i^2}{2E}
\end{equation}:

\section*{Appendix 4: Derivation of equation (\ref{eq::pol_osc_formula})}
\begin{eqnarray*} 
P({`` \nue `` } \rightarrow {`` \numu ``}) \hspace{0.7\textwidth} \mbox{} 
\end{eqnarray*}
\begin{eqnarray}
=& 
\left| 
  \left( 
    \begin{array}{c}
       0 \\ 1 
    \end{array} 
  \right) 
  U 
  \left(
     \begin{array}{cc} 
        e^{- i 2\pi L n_{\rm ord} /\lambda} & 0 \\ 0 & e^{- i 2 \pi L n_{\rm extra} / \lambda } 
     \end{array}
  \right)
  U^{-1}
  \left( 
    \begin{array}{c} 
       1 \\ 0 
    \end{array} 
  \right) 
\right| ^2  \\
=& 
\left| 
  \left( 
    \begin{array}{c}
       0 \\ 1 
    \end{array} 
  \right) 
  \left( 
    \begin{array}{cc} 
      \costh & -\sinth \\ \sinth & \costh 
    \end{array} 
  \right) 
  \left(
     \begin{array}{cc} 
        e^{-i 2\pi n_{\rm ord} L /\lambda} & 0 \\ 0 & e^{-i 2 \pi  n_{\rm extra} L / \lambda } 
     \end{array}
   \right)
  \left(
    \begin{array}{cc} 
       \costh & \sinth \\ -\sinth & \costh
    \end{array}
  \right) 
  \left( 
    \begin{array}{c} 
       1 \\ 0 
    \end{array} 
  \right) 
\right| ^2  \\
=& 
\left| 
  \left( 
    \begin{array}{c}
       0 \\ 1 
    \end{array} 
  \right) 
  \left( 
    \begin{array}{cc} 
      \costh & -\sinth \\ \sinth & \costh 
    \end{array} 
  \right) 
 e^{-i 2\pi n_{\rm ord} L /\lambda}
  \left( 
     \begin{array}{cc} 
        1 & 0 \\ 0 & e^{-i 2 \pi  (n_{\rm extra}-n_{\rm ord}) L / \lambda } 
      \end{array} 
  \right) 
  \left( 
    \begin{array}{c} 
      \costh \\ -\sinth 
    \end{array} 
  \right) 
\right| ^2 \\
=& 
\left| 
  \left( 
    \begin{array}{c}
       0 \\ 1 
    \end{array} 
  \right) 
  \left( 
    \begin{array}{cc} 
      \costh & -\sinth \\ \sinth & \costh 
    \end{array} 
  \right) 
  \left( 
    \begin{array}{cc} 
        1 & 0 \\ 0 & e^{-i 2 \pi  \Delta n L/ \lambda } 
    \end{array} 
  \right)
  \left( 
    \begin{array}{c} 
      \costh \\ -\sinth 
    \end{array} 
  \right) 
\right| ^2 \label{eq::pol2_deltan}\\
=& 
\sin^2 (2\theta) \sin^2 \left(\frac{\Delta n L}{\lambda / \pi} \right) \label{eq::pol2_osc_formula}\\
= & \sin^2 (2\theta) \sin^2 \left( \pi \frac{L}{\lambda_{\rm osc} } \right) 
\quad {\rm with} \quad \lambda_{\rm osc} = \frac{\lambda}{\Delta n} \label{eq::pol2_osc_length}
\end{eqnarray}
We have abbreviated the calculation between equations (\ref{eq::pol2_deltan}) and (\ref{eq::pol2_osc_formula}) since there are exactly the
same transformation steps as between equations (\ref{eq::deltam2}) and (\ref{eq::osc2_formula}).

\end{document}